\documentclass[aip,jap, reprint,amssymb,amsmath,fleqn]{revtex4-1}

\usepackage{docs}%
\usepackage{bm}%
\usepackage[colorlinks=true,linkcolor=blue]{hyperref}%
\usepackage{graphicx}
\expandafter\ifx\csname package@font\endcsname\relax\else
 \expandafter\expandafter
 \expandafter\usepackage
 \expandafter\expandafter
 \expandafter{\csname package@font\endcsname}%
\fi
\begin{document}

\title{Effects of trace elements on the crystal field parameters of Nd ions at the surface of Nd$_2$Fe$_{14}$B grains}%

%
%\author{Yuta Toga, 
%Tsuneaki Suzuki, and 
%Akimasa Sakuma}%
%\email{sakuma@solid.apph.tohoku.ac.jp}
%\affiliation{Department of Applied Physics, Tohoku University, Sendai 980-8579, Japan}%
%

\author{Yuta Toga}
\author{Tsuneaki Suzuki}%
\author{Akimasa Sakuma}
\email{sakuma@solid.apph.tohoku.ac.jp}
\affiliation{%
Department of Applied Physics, Tohoku University, Sendai 980-8579, Japan%\\This line break forced% with \\
}%

%\date{}%
%\revised{}%

\begin{abstract}
Using first-principles calculations,
we investigate the positional dependence of trace elements 
such as O and Cu on the crystal field parameter $A_2^0$, proportional to the magnetic anisotropy constant $K_u$
of Nd ions placed at the surface of Nd$_2$Fe$_{14}$B grains.
The results suggest the possibility that the $A_2^0$ parameter of Nd ions at the (001) surface
of Nd$_2$Fe$_{14}$B grains exhibits a negative value when the O or Cu atom is located near the surface, 
closer than its equilibrium position.
At the (110) surface, however, O atoms located at the equilibrium position provide a negative $A_2^0$, 
while for Cu additions $A_2^0$ remains positive regardless of Cu's position.
Thus, Cu atoms are expected to maintain a positive local $K_u$ of surface Nd ions
more frequently than O atoms when they approach the grain surfaces in the Nd-Fe-B grains.
\end{abstract}

\maketitle

\section{INTRODUCTION}
Rapidly increasing demand for efficient electric motors motivates the development of high-performance Nd-Fe-B magnets. 
In sintered Nd-Fe-B magnets, Nd is frequently substituted with Dy, owing to Dy's suppression of coercivity ($H_c$) degradation.
However, Dy is expensive and decreases the magnetization of Nd-Fe-B magnets.
To realize Dy-free high-performance Nd-Fe-B magnets, 
we must understand the $H_c$ mechanism of rare-earth (Re) permanent magnets.
Although the $H_c$ mechanism has been discussed for sintered Nd-Fe-B magnets,\cite{r1,r2,r3}
our understanding of $H_c$ is incomplete, requiring further examination.

Recent papers emphasized that the development of high-coercivity Nd-Fe-B magnets requires understanding their microstructure, especially the grain boundary (GB) phase surrounding the Nd$_2$Fe$_{14}$B grains.\cite{r4,r5} 
Researchers have reported that the intergranular Nd-rich phase includes neodymium oxides (NdO$_x$) with diverse crystal structures (e.g., fcc, hcp), suggesting that O atoms must exist near Nd atoms at the interface between the Nd-rich phase and Nd$_2$Fe$_{14}$B grains.\cite{r6,r7}
Recently, Amin {\it et al.}\cite{r8} confirmed the segregation of Cu to the NdO$_x$/Nd$_2$Fe$_{14}$B interface after annealing, suggesting that the magnetic anisotropy constant $K_u$ grains are lapped by a Cu-rich layer.
Thus, O and Cu atoms are considered to stay around the Nd ions at the interfaces,
playing a key role for the coercive force of Nd-Fe-B sintered magnets.

%
%{\color{red}
From theoretical viewpoint, it is widely accepted that the magnetic anisotropy of rare earth magnets is mainly dominated by the 4f electrons in the rare earth ions, because of their strongly anisotropic distribution due to their strong intra-atomic interactions such as electron-correlation and $L$-$S$ coupling.
The one-electron treatments based on the local density functional approximation have not yet been successfully reproducing this feature in a quantitative level, in contrast to the 3d electronic systems.
To overcome this problem, crystalline electric field theory combined with the atomic many-body theory has been adopted by many workers to study the magnetic anisotropy.
In 1988, Yamada {\it et al.}\cite{yamada} successfully reproduced magnetization curves reflecting magnetic anisotropy of the series of Re$_2$Fe$_{14}$B, using the crystal field parameters $A_l^m$ as adjustable parameters.
In 1992, the first principles calculations to obtain the $A_l^m$ were performed by Richter {\it et al.}\cite{a20_1} for ReCo$_5$ system and F\"{a}hnle {\it et al.}\cite{a20_2} for Re$_2$Fe$_{14}$B system.

Based on this concept, Moriya {\it et al.}\cite{r9} showed via first-principles calculations that the crystal field parameter $A_2^0$ on the Nd ion exhibits a negative value when the Nd ion in the (001) planes is exposed to vacuum.
As shown by Yamada {\it et al.}\cite{yamada}, the magnetic anisotropy constant $K_u$ originated from rare earth ion is approximately proportional to $A_2^0$ when the exchange field acting on the 4f electrons in a rare earth ion is sufficiently strong.
Since the proportional coefficient is positive for Nd ion, negative $A_2^0$ implies the planar magnetic anisotropy.
%}
%
%
%
%From a theoretical viewpoint, Moriya {\it et al.}\cite{r9} showed via first-principles calculations that the crystal field parameter $A_2^0$ on the Nd ion exhibits a negative value when the Nd ion in the (001) planes is exposed to vacuum.
%It is widely accepted that the anisotropy constant $K_u$ is approximately proportional to $A_2^0$ when the exchange field acting on the 4f electrons in a rare earth ion is sufficiently strong.
Furthermore, Mitsumata {\it et al.}\cite{r10} demonstrated that a single surface atomic layer with a negative $K_u$ value dramatically decreases $H_c$.
In physical sintered magnets, however, the grain surfaces are not exposed to vacuum but instead face GB phases.
Therefore, our next step is studying how interface elements in the GB phases adjacent to Nd ions affect the $A_2^0$ values at the Nd-Fe-B grain surface.
However, note that, in an actual system, many atoms interact with surface Nd ions and many possible configurations exist near the interface between GB and Nd$_2$Fe$_{14}$B phases.  
In this case, it is important from a theoretical standpoint to provide separate information on the positional ($r, \theta$) dependence of individual atoms on the local $K_u$ (i.e., $A_2^0$) of Nd ions at the grain surface.
These analyses may provide useful information to judge the factors dominating the coercive force during experimental observation of atomic configurations in real systems in the future.

In this study, we investigate the influence of O and Cu atoms on the $A_2^0$ of the Nd ion placed at the surface of Nd$_2$Fe$_{14}$B grains via first-principles calculations.
As the sign of the single-site $K_u$ of an Nd ion is the same as that of $A_2^0$, the evaluation of $A_2^0$ is useful to understand how trace elements affect $K_u$ at the surface of Nd${_2}$Fe${_{14}}$B.
We select O and Cu as trace elements on the basis of experimental results.

%
%\,{\AA}
\section{COMPUTATIONAL DETAILS}
%
% Fig 1 %%%%%%%%%%%%%%%%%%%%%%%%%%%%%%%%%%%%%%%%%%
\begin{figure}
\includegraphics[width=8.5cm,clip]{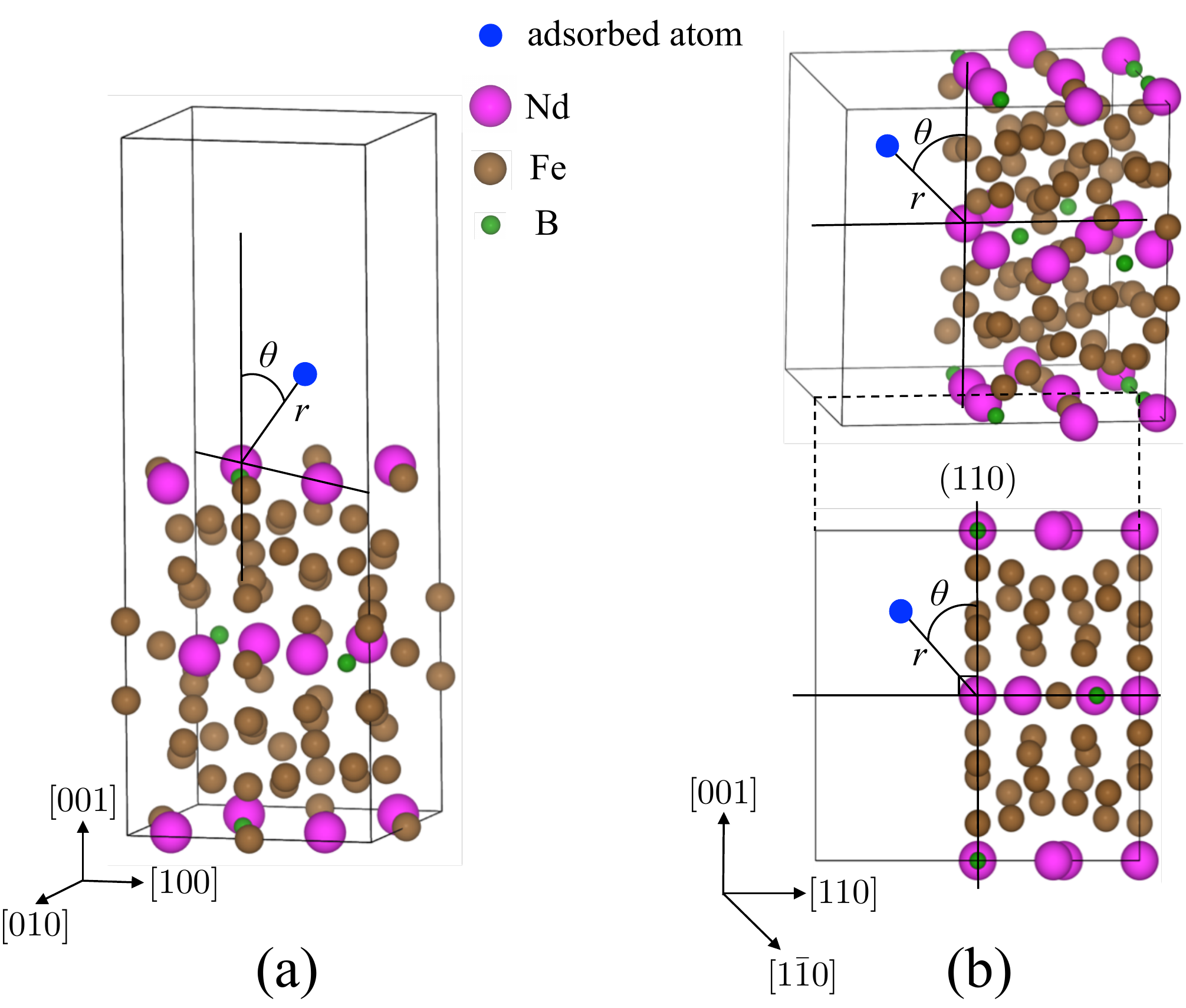}% Here is how to import EPS art
\caption{\label{f1}The geometric relationship between the Nd ion on the surface of Nd$_2$Fe$_{14}$B and the trace element for the (a)\,(001) and (b)\,(110) surface slab models.
Here, $r$ indicates the distance between the Nd ion and the trace element, and $\theta$ indicates the angle between the c-axis of Nd$_2$Fe$_{14}$B and the direction of $r$.
This figure is plotted by using VESTA.\cite{vesta}}
%
%Unit cell of the slab model for the (a)\,(001) and (b)\,(110) surface of Nd$_2$Fe$_{14}$B.
%The (001) slab model has a vacuum layer with a thickness of 12.19\,{\AA} along the c-axis, and the (110) slab model has a vacuum layer with a thickness of 6.22\,{\AA} along the a-axis. In the (110) slab model, the (110) direction in the original unit cell corresponds to the a-axis.\\
%\\
%Schematic of the geometric relationship between the Nd ion on the surface of Nd$_2$Fe$_{14}$B and the trace element for the (a)\,(001) and (b)\,(110) slab models. Here, $r$ indicates the distance between the Nd ion and the trace element, and θ indicates the angle between the c-axis of Nd$_2$Fe$_{14}$B and the direction of $r$.
%In the (110) slab model, the (110) direction in the original unit cell corresponds to the a-axis.
\end{figure}
%%%%%%%%%%%%%%%%%%%%%%%%%%%%%%%%%%%%%%%%%%%
%
Electronic structure calculations were performed by density functional theory using the Vienna ab initio simulation package (VASP 4.6).
%{\color{red}The ionic potentials are described by projector-augmented-wave (PAW) method and the exchange-correlation energy of the electrons is described within a generalized gradient approximation (GGA).\cite{r11}}
%{\color{red}
The 4f electrons in the Nd ions were treated as core electrons in the electronic structure calculations for the valence electrons, based on the concept mentioned in the previous section.
%}
%{\color{red}
The ionic potentials are described by projection-augmented-wave (PAW) method\cite{paw} and the exchange-correlation energy of the electrons is described within a generalized gradient approximation (GGA).
We used the exchange-correlation function determined by Ceperly and Alder and parametrized by Perdew and Zunger.\cite{GGA}
%}

We examined the (001) and (110) surfaces of Nd$_2$Fe$_{14}$B with the addition of the trace element using slab models.
Figure\,\ref{f1} shows the geometric relationship between the Nd ion at the surface of Nd$_2$Fe$_{14}$B and the trace element for the (001) and (110) surface models.  We placed the trace element at various distances $r$ around the Nd ion at the surface, and at various angles $\theta$ between the c-axis of Nd$_2$Fe$_{14}$B and the direction of $r$.
In the (001) surface model (Fig.\,\ref{f1}(a)), this unit cell has a vacuum layer equivalent to the thickness of the Nd$_2$Fe$_{14}$B unit cell along the c-axis (12.19\,{\AA}) and consists of 12 Nd, 58 Fe, and six B atoms.
In the (110) surface model (Fig.\,\ref{f1}(b)), we restructured the Nd$_2$Fe$_{14}$B unit cell to expose Nd atoms on the (110) surface.
The (110) direction in the original unit cell corresponds to the a-axis in the restructured unit cell.
The restructured unit cell consists of 12 Nd, 68 Fe, and six B atoms.
%{\color{red}
The lattice constant of the a-axis parallel to the (110) surface is $\sqrt{2}$ and perpendicular to it is $1/\sqrt{2}$ times that of the original Nd$_2$Fe$_{14}$B unit cell.
%}

The (110) surface model has a vacuum layer with the thickness of $8.8/\sqrt{2}=6.22$\,{\AA} along the (110) direction.
The mesh of the numerical integration was provided by a discrete Monkhorst-Pack $k$-point sampling.

To investigate the magnetic anisotropy of this system, we calculated $A_2^0$ for Nd ion adjacent to a trace element at the Nd$_2$Fe$_{14}$B surface.
The value of the magnetic anisotropy constant $K_u$ originated from rare earth ion is approximately given by $K_u=-3J(J-1/2)\alpha\langle r^2 \rangle A_2^0$ when the exchange field acting on the 4f electrons in a rare earth ion is sufficiently strong.
Here, $J$ is the angular momentum of 4f electronic system, $\alpha$ means the Stevens factor characterizing the rare earth ion, and $\langle r^2 \rangle$ is the spatial average of $r^2$, as given by Eq.\,\eqref{eq3} below.
For Nd ion, $J=9/2$ and $\alpha$ is negative, and then positive $A_2^0$ leads to positive $K_u$.  The physical role of $A_2^0$ is to reflect the electric field from surrounding charge distribution acting on the 4f electrons whose spatial distribution differs from spherical one due to the strong $L$-$S$ coupling.
Therefore, to obtain the value of $A_2^0$, one needs the charge distribution surrounding the 4f electronic system with the following equation:\cite{a20_1,a20_2}
%}

%We calculated $A_2^0$ for an Nd ion adjacent to a trace element at the Nd$_2$Fe$_{14}$B surface.
%The value of $A_2^0$ was calculated with the following equations:\cite{a20_1,a20_2}
%

\begin{eqnarray}
A_2^0&=&-\frac{e}{4\pi\epsilon_0}\frac{4\pi a}{5}
\int d\bm{R} \rho(\bm{R})Z_2^0(\bm{R})\nonumber\\&& \times
\int dr r^2\frac{r_<^2}{r_>^3}4\pi\rho_{4f}(r)/\langle r^2\rangle,
\label{eq1}\\
Z_2^0 (\bm{R})&=&a(3R_z^2-|\bm{R}|^2)/|\bm{R}|^2,
\label{eq2}\\
\langle r^2 \rangle &=&\int dr r^2 4\pi\rho_{4f}(r)r^2,
\label{eq3}
\end{eqnarray}
where, $a=(1/4)(5/\pi)^{1/2}$, $r_<=\min(r,|\bm{R}|)$, and $r_>=\max(r,|\bm{R}|)$.
Here, $\rho_{4f}(r)$ is the radial part of the 4f electron probability density in the Nd ion, 
and $\rho(\bm{R})$ is the charge density including the nuclei and electrons.
The integral range of $\bm{R}$ in Eq.\,\eqref{eq1} is within a sphere with a radius of 70{\AA} from the surface Nd site.
Here, the valence electron density was calculated by VASP 4.6, and $\rho_{4f}(r)$ was calculated using an isolated Nd atom.
In $\rho(\bm{R})$, the core electrons (including the 4f electron) forming pseudo-potentials in VASP were treated as point charges as well as nuclei.

%{\color{red}
%We should mention the integral range of $\bm{R}$ in Eq.\,\eqref{eq1}.
%Figure\,\ref{fr1} shows the $A_2^0$ of Nd ion for the surface models without trace element as a proportional to the integral range.
%%
%We can see sudden jumps of $A_2^0$ which are caused by the nuclei and the core electrons.
%Especially in the region of $R_c<20$\,{\AA}, since deviations from electrical neutrality are large, $A_2^0$ fluctuates sensitively depending on $R_c$.
%In the Nd$_2$Fe$_{14}$B, it is known that the Nd ion's own valence charge make a dominant contribution to $A_2^0$.\cite{a20_2}
%In Fig.\,\ref{fr1}, indeed, values of $A_2^0$ for $R_c\sim 1.7$\,{\AA} (corresponding to Nd atomic sphere radius) are qualitatively consistent with convergence values.
%However, $A_2^0$ depends very sensitively on the choice of the Nd atomic sphere radius if trace element is placed in close proximity to the Nd ion.
%Therefore, we adopt, as the integral range of $\bm{R}$ in Eq.\,\eqref{eq1}, within a sphere with a radius of 70\,{\AA} from the surface Nd site. }
%
%
% Fig r1 %%%%%%%%%%%%%%%%%%%%%%%%%%%%%%%%%%%%%%%%%%
%%
%\begin{figure}
%\includegraphics[width=7.8cm,clip]{fig/r01_a20dis}% Here is how to import EPS art
%\caption{\label{fr1} Contribution to the $A_2^0$ of the Nd ion from charge $\rho(\bm{R})$ within a sphere with a radius of $R_c$ from the Nd site for the (001) and the (110) surface model without trace element. }
%
%\end{figure}
%%
%%%%%%%%%%%%%%%%%%%%%%%%%%%%%%%%%%%%%%%%

\section{RESULTS AND DISCUSSION}
As shown in a previous paper,\cite{r12} our computational procedure produces $A_2^0$ values of bulk Nd$_2$Fe$_{14}$B
reasonably consistent with those obtained using the full-potential linear-muffin-tin-orbital method.\cite{r13}
In addition, our calculated $A_2^0$ for the Nd ion at the (001) surface of Nd$_2$Fe$_{14}$B exhibits a negative value around $-300$\,$\mathrm{K}/a_B^2$ where $a_B$ represents the Bohr radius.
This corresponds to that calculated using the full-potential linearized augmented plane wave plus local orbitals (APW+lo) method implemented in the WIEN2k code.\cite{r9}
Therefore, our numerical calculation methods described in section II should be sufficiently reliable to quantitatively evaluate the influence of a trace element on $A_2^0$ acting on the Nd ion at the surface of Nd$_2$Fe$_{14}$B.

%%
%% Fig 2 %%%%%%%%%%%%%%%%%%%%%%%%%%%%%%%%%%%%%%%%%%
%%%
%\begin{figure}
%\includegraphics[width=7.8cm,clip]{fig/2.pdf}% Here is how to import EPS art
%\caption{\label{f2}Schematic of the geometric relationship between the Nd ion on the surface of Nd$_2$Fe$_{14}$B and the trace element for the (a)\,(001) and (b)\,(110) slab models. Here, $r$ indicates the distance between the Nd ion and the trace element, and θ indicates the angle between the c-axis of Nd$_2$Fe$_{14}$B and the direction of $r$.
%In the (110) slab model, the (110) direction in the original unit cell corresponds to the a-axis.}
%\end{figure}
%%%
%%%%%%%%%%%%%%%%%%%%%%%%%%%%%%%%%%%%%%%%%
%
%
%{\color{blue} \sout{Figure \ref{f2} shows the geometric relationship between the Nd ion at the surface of Nd$_2$Fe$_{14}$B and the trace element for the (001) and (110) surface models.  We placed the trace element at various distances $r$ around an Nd ion at the surface, and at various angles $\theta$ between the c-axis of Nd$_2$Fe$_{14}$B and the direction of $r$.}}

% Fig 3 %%%%%%%%%%%%%%%%%%%%%%%%%%%%%%%%%%%%%%%%%%%%%%%
\begin{figure}
\includegraphics[width=8cm,clip]{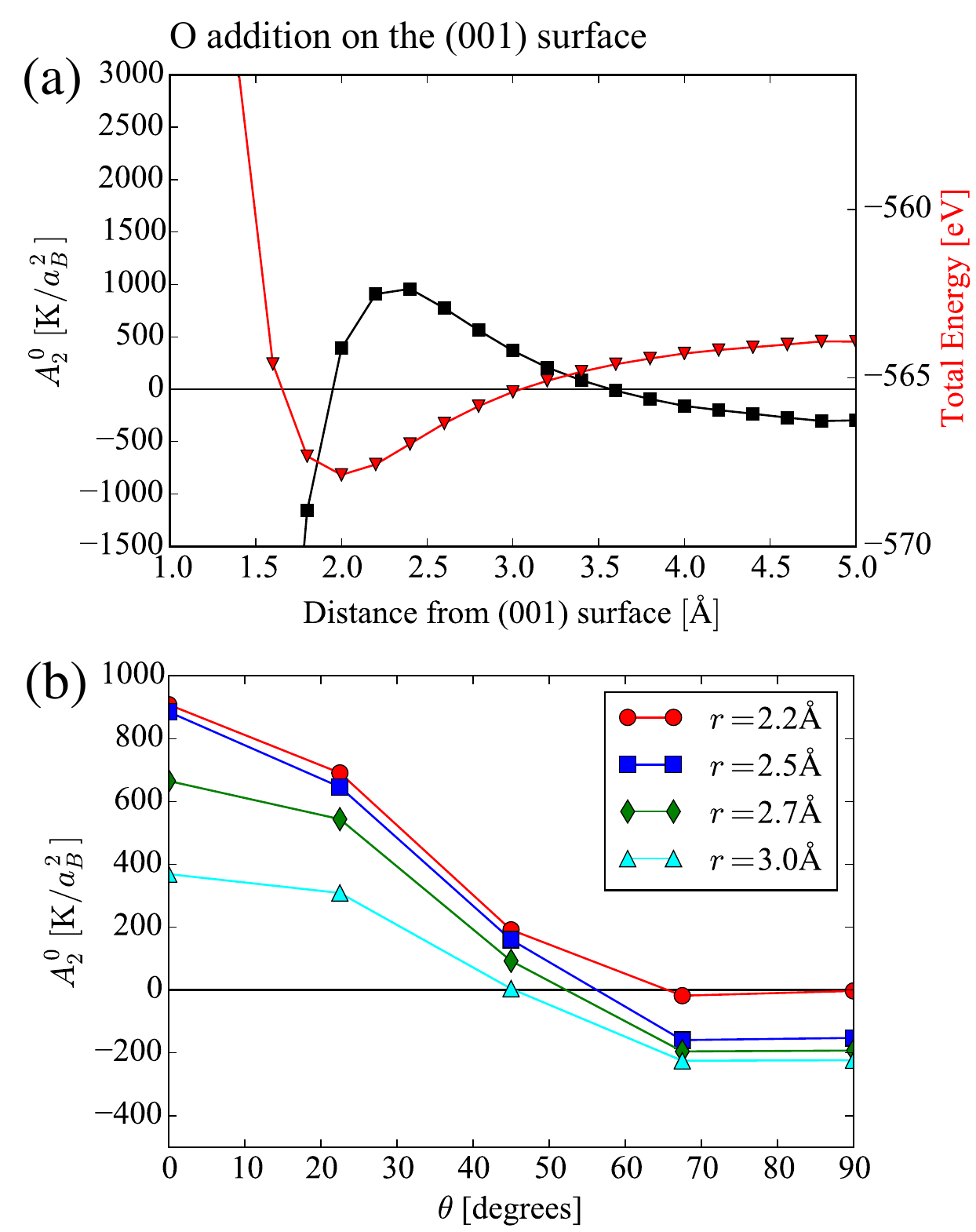}% Here is how to import EPS art
\caption{\label{f3}(a)\,The $r$ dependence of crystal field parameter $A_2^0$ and the total energy for O addition on the (001) surface of Nd$_2$Fe$_{14}$B for $\theta=0^\circ$.
The closed squares and triangles indicate $A_2^0$ and total energies, respectively.
(b)\,The $\theta$ dependence of $A_2^0$ for $r=$ 2.2\,{\AA}, 2.5\,{\AA}, 2.7\,{\AA}, and 3.0\,{\AA}.
}
\end{figure}
%%%%%%%%%%%%%%%%%%%%%%%%%%%%%%%%%%%%%%%%
%
%
%
Figure \ref{f3}(a) shows variations in $A_2^0$ and electronic total energy when the O atom approaches the (001) surface Nd ion with angle $\theta=0^\circ$ (see Fig.\,\ref{f1}(a)).
We confirm, as predicted by the previous work,\cite{r9} that $A_2^0$ exhibits a negative value when O is positioned at $r>$ 3.5\,{\AA} from the surface.
%{\color{red}
At this situation, the total energy is confirmed to be almost equal to the summation of separated system of (001)-slab and O atom, each of which is -561.838 and -1.826\,eV. Thus the deviation energies from the summation of these values represent the interaction energies between the slab and O atom.
%}
When O nears the Nd ion at $\theta=0^\circ$, $A_2^0$ becomes positive and increases to a peak value around 1000\,$\mathrm{K}/a_B^2$ at $r=$ 2.4\,{\AA}.
Interestingly, with further decrease of $r$, $A_2^0$ abruptly drops into negative values.

% Fig 4 %%%%%%%%%%%%%%%%%%%%%%%%%%%%%%%%%%%%%%%%%%%%%%%
\begin{figure}
\includegraphics[width=8.5cm,clip]{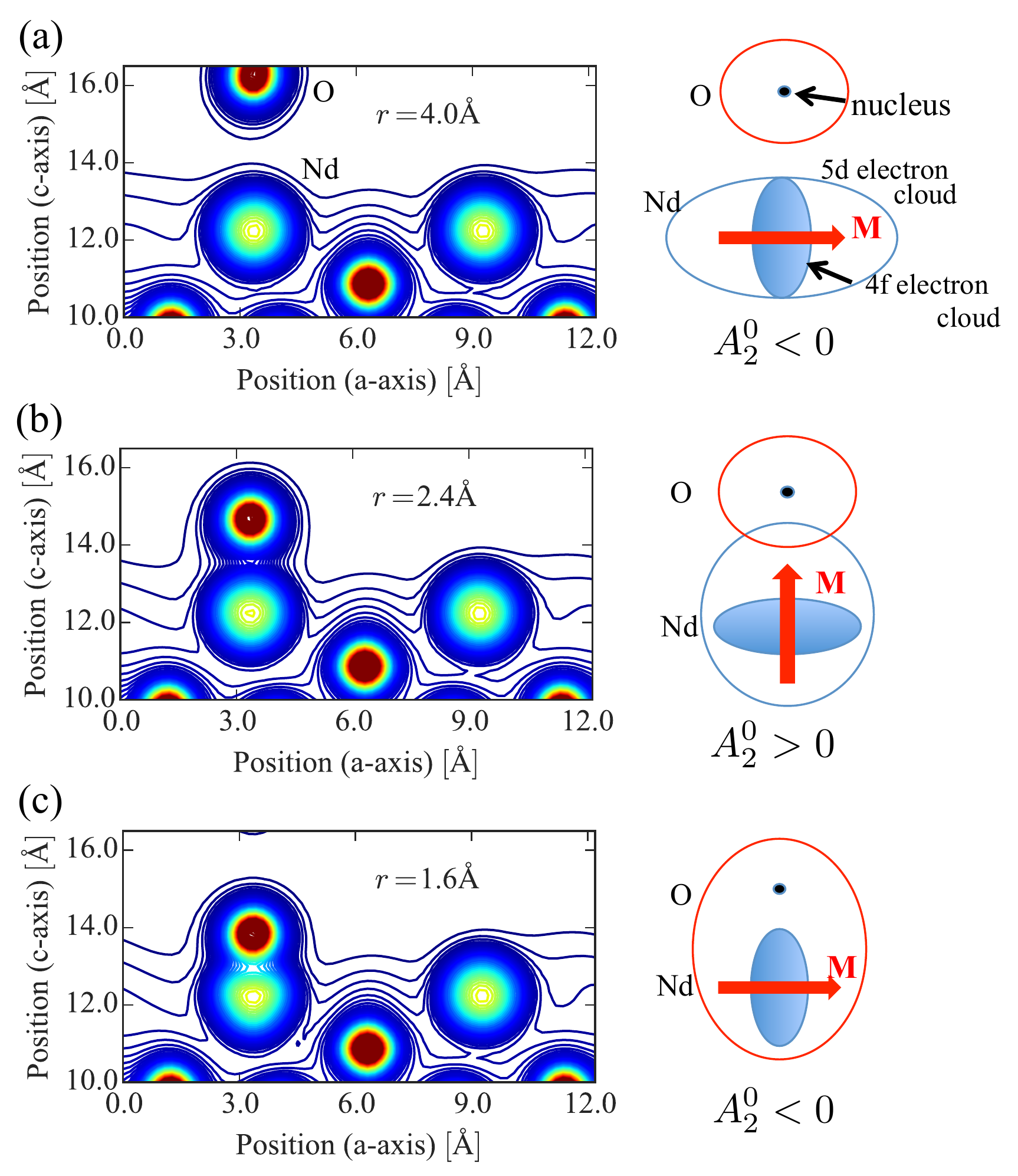}% Here is how to import EPS art
\caption{\label{f4}Calculated valence electron density distributions for O addition on the (001) surface for $r=$ (a)\,4.0\,{\AA}, (b)\,2.4\,{\AA}, and (c)\,1.6\,{\AA}, together with schematics corresponding to these three cases.}
\end{figure}
%%%%%%%%%%%%%%%%%%%%%%%%%%%%%%%%%%%%%%%%
%
%
To understand these behaviors, we show in Fig.\,\ref{f4} the calculated electron density distributions for $r=$ (a)\,4.0\,{\AA}, (b)\,2.4\,{\AA}, and (c)\,1.8\,{\AA}, together with schematics corresponding to these three cases.
The increase of $A_2^0$ with decreasing $r$ for $r>2.4$\,{\AA} can be explained by the change in 5d electron distribution surrounding the 4f electrons of the Nd ion.
That is, the 5d electron cloud extends towards the O atom through hybridization (Fig.\,\ref{f4}(b)), which repositions the 4f electron cloud within the c-plane in order to avoid the repulsive force from the 5d electrons; this produces positive values of $A_2^0$.
The decreasing behavior for $r<2.2$\,{\AA} is attributed to the influence of the positively charged nucleus in the O atom exceeding the hybridization effect of the valence electrons (Fig.\,\ref{f4}(c)); the attractive Coulomb force from O nucleus could rotate the 4f electron cloud so as to minimize the electro-static energy, resulting in negative values of $A_2^0$.

The variation of electronic total energy in Fig.\,\ref{f3}(a) indicates a stable distance around $r=$ 2.0\,{\AA}, at which the value of $A_2^0$ is still positive.
However, the value of $A_2^0$ easily becomes negative with only a 0.2\,{\AA} decrease from the equilibrium position, with an energy cost less than 1\,eV.
This deviation can take place due to stresses, defects, or deformations around grain boundaries in an actual system.
This implies that the value of $A_2^0$ may exhibit a negative value at real grain surfaces adjacent to GB phases, rather than to vacuum space.

In Fig.\,\ref{f3}(b), we show the $\theta$-dependences of $A_2^0$ for various values of $r$.
Starting from positive values at $\theta=0^\circ$, $A_2^0$ decreases monotonically with increasing $\theta$ and reaches negative values for $\theta>50^\circ$.
It is meaningful to compare these behaviors with those of the point charge model.
In the point charge model, $A_2^0$ is proportional to $Z_2^0\propto –eq(3\cos\theta^2-1)\ (e>0)$ where $q$ is the valence number of the point charge.
Thus the gross features of these results can be realized by the point charge model, if one assumes a negative charge ($q<0$) for the O ion.
Note, however, that we do not clearly identify the negative charge on the O atom when $r\geq 2.2$\,{\AA}.
Therefore, the oxygen atom should be considered to influence the redistribution of valence electrons within the Nd atomic sphere such that the point charge model is applicable as if O were negatively charged.
This can be regarded as a sort of screening effect.\cite{r13_2}
Similar explanations were proposed for the N effects in Re$_2$Fe$_{17}$N$_3$\cite{r14} and ReFe$_{12}$N,\cite{r15,r16} systems, where the N atoms changed the magnetocrystalline anisotropy energies of these systems.

% Fig 5 %%%%%%%%%%%%%%%%%%%%%%%%%%%%%%%%%%%%%%%%%%%%%%%%%%
\begin{figure}
\includegraphics[width=8cm,clip]{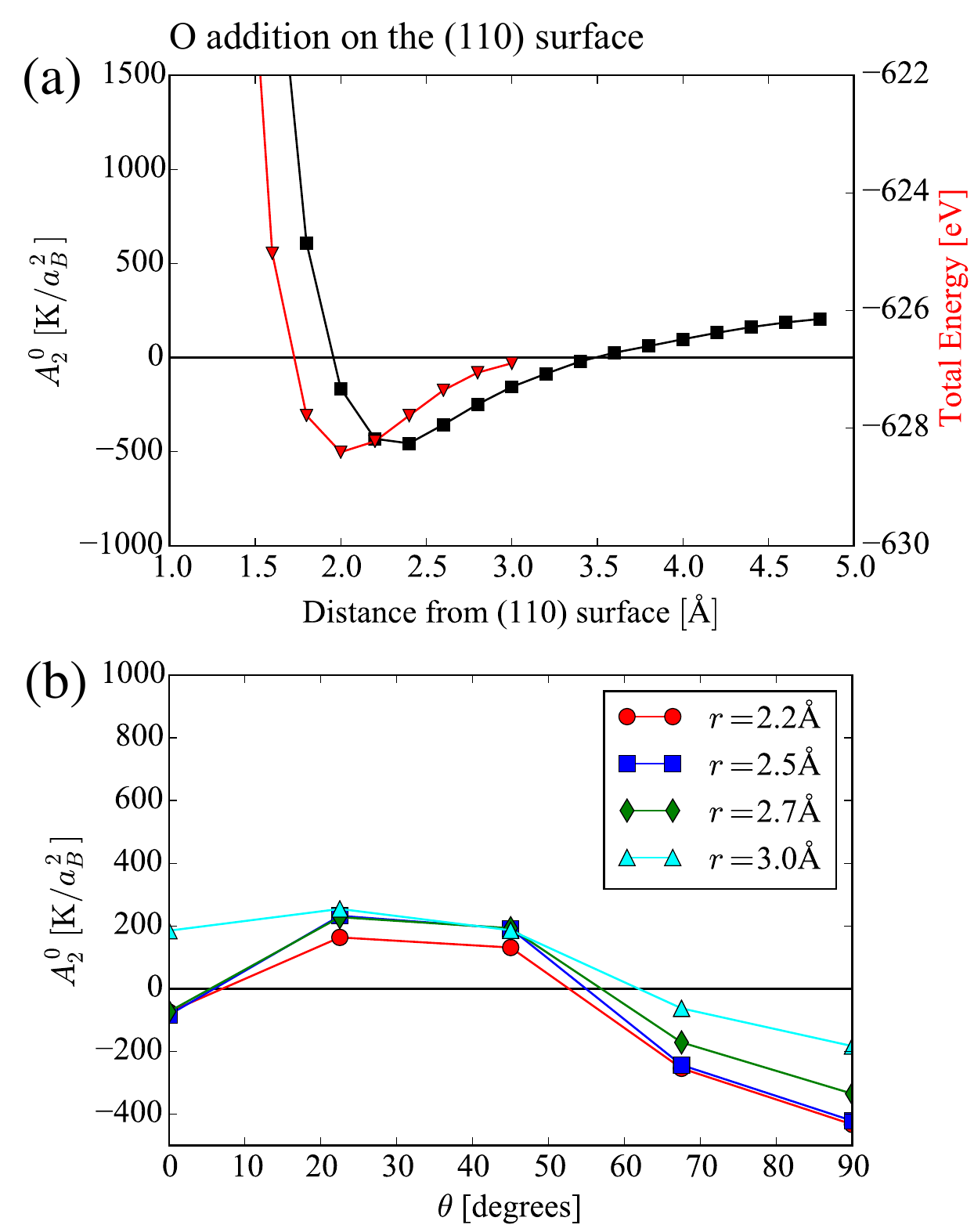}% Here is how to import EPS art
\caption{\label{f5}(a)\,The $r$ dependence of crystal field parameter $A_2^0$ and the total energy for O addition on the (110) surface of Nd$_2$Fe$_{14}$B for $\theta=90^\circ$.
The closed squares and triangles indicate $A_2^0$ and total energies, respectively.
(b)\,$\theta$ dependence of $A_2^0$ for $r=$ 2.2\,{\AA}, 2.5\,{\AA}, 2.7\,{\AA}, and 3.0\,{\AA}.}
\end{figure}
%%%%%%%%%%%%%%%%%%%%%%%%%%%%%%%%%%%%%%%%%%%%
%
%
Figure \ref{f5}(a) shows the $r$-dependence of the values of $A_2^0$ and electronic total energy when the O atom approaches the Nd ion in the (110) surface with an angle $\theta=90^\circ$ from the c-axis.
Contrary to the case for the (001) surface, $A_2^0$ is positive when Nd ions in the (110) surface are exposed to vacuum, and when the O atom is far from the surface.
%{\color{red}
In this case, since the thickness of the vacuum space is only 6.22\,{\AA}, the total energy exhibits symmetric variation with respect to r=3.1\,{\AA}.  From this reason, we present total energies within the range $r\leq 3.1$\,{\AA}.
%}
When the O atom approaches the Nd ion, the value of $A_2^0$ becomes negative for $r<$ 3.5\,{\AA} and stabilizes around $r=2.0$\,{\AA}, where the value remains negative.  Notably, the value of $A_2^0$ becomes positive with only a 0.2\,{\AA} decrease in $r$.
This is due to the attraction of the O nucleus to the 4f electron cloud, which aligns the 4f moment with the direction of the c-axis.
The energy consumption for this reduction of $r$ is less than 1\,eV.
In Fig.\,\ref{f5}(b), we show the $\theta$-dependence of the values of $A_2^0$ for varying $r$.
As shown in Fig.\,\ref{f5}(a), the values of $A_2^0$ for $\theta=90^\circ$ are negative in the range of 2.2\,{\AA}$<r<$3.0\,{\AA}.
Similarly to the (001) surface in Fig.\,\ref{f3}(b), $A_2^0$ increases with decreasing $\theta$.
This can also be understood from the O atom's redistribution of the valence electrons within the Nd atomic sphere, as if a negative ion exists in the direction of the O atom.
The reason for the slight decrease in $A_2^0$ at $\theta=0^\circ$ is not clear at this stage.

% Fig 6 %%%%%%%%%%%%%%%%%%%%%%%%%%%%%%%%%%%%%%%%%%%%%%%%%%%%%
\begin{figure}
\includegraphics[width=8cm,clip]{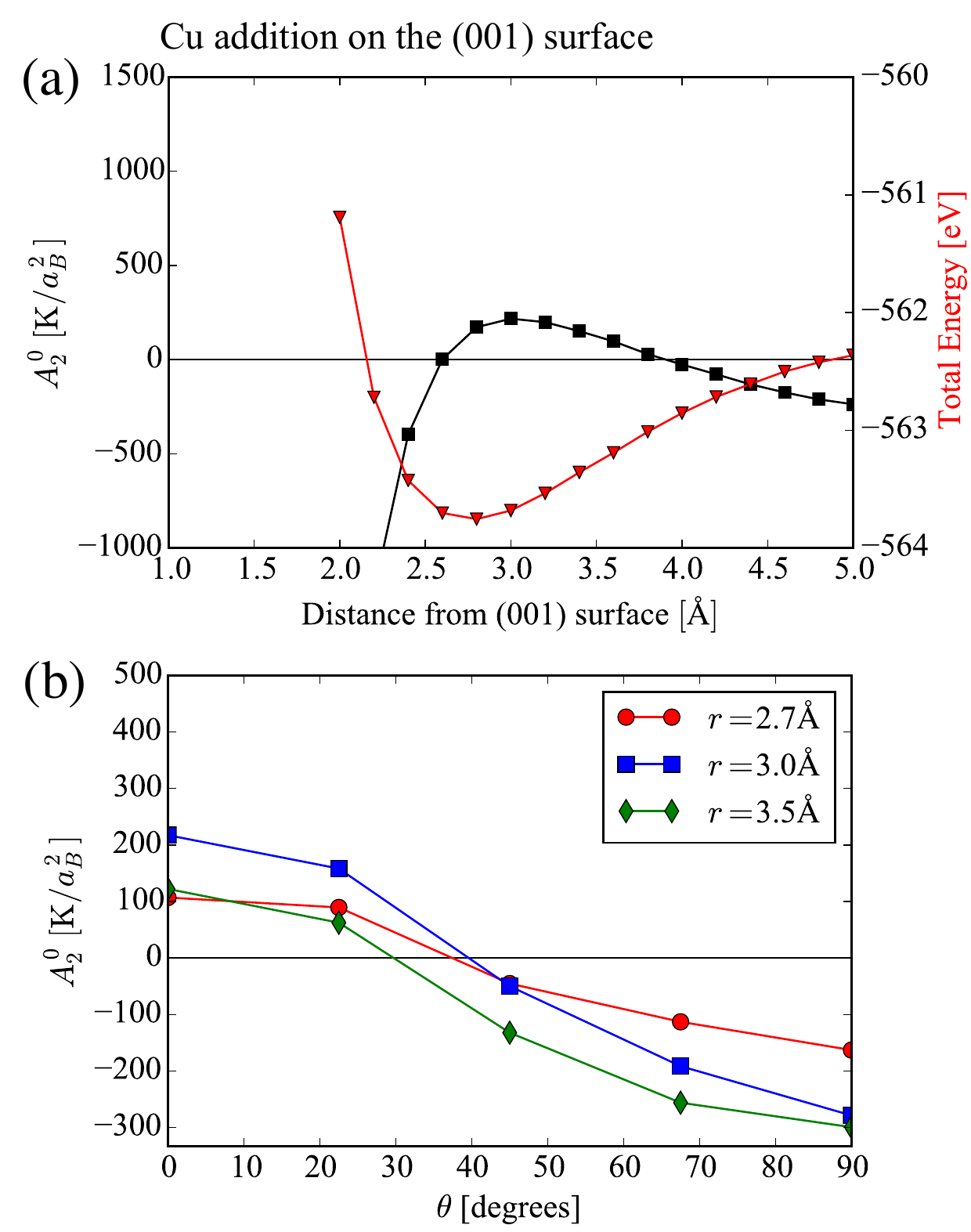}% Here is how to import EPS art
\caption{\label{f6}(a)\,The $r$ dependence of crystal field parameter $A_2^0$ and the total energy for Cu addition on the (001) surface of Nd$_2$Fe$_{14}$B for $\theta=0^\circ$. 
The closed squares and triangles indicate $A_2^0$ and total energies, respectively.
(b)\,The $\theta$ dependence of $A_2^0$ for $r=$ 2.7\,{\AA}, 3.0\,{\AA}, 3.5\,{\AA}.}
\end{figure}
%%%%%%%%%%%%%%%%%%%%%%%%%%%%%%%%%%%%%%%%%%%%%
%
%
Next we proceed to the case of Cu addition.
Figure\,\ref{f6}(a) shows the $r$-dependence of the values of $A_2^0$ and total energy for Cu addition with $\theta=0^\circ$.
The behaviors of $A_2^0$ and total energy are almost the same as those for O addition, shown in Fig.\,\ref{f3}(a).
However, the variations are less dramatic compared to the O case.
This may reflect the weaker hybridization of the Nd atom with Cu than with O.
The peak position of $r$ (around 3.0\,{\AA}) is greater than that with O addition, which may be due to the large atomic radius of Cu.
In this case, a decrease of about 0.5\,{\AA} in $r$ from the equilibrium position causes $A_2^0$ to become negative.
This move costs around 0.3\,eV in energy, which is less than that for O addition.
The $\theta$-dependence of $A_2^0$ for various values of $r$ is shown in Fig.\,\ref{f6}(b).
%Except for the case for $r=$ 2.5\,{\AA},
The behavior can be explained through the hybridization between valence electrons of both Nd and Cu atoms, as in the case for O addition.
%The value of $A_2^0$ at $r=$ 2.5\,{\AA} is almost insensitive to variations in $\theta$.
%This may be due to the balancing of the redistribution of valence electrons in the Nd ion with the positive charge of the nucleus of the Cu atom.
\\
% Fig 7 %%%%%%%%%%%%%%%%%%%%%%%%%%%%%%%%%%%%%%%%%%%%%%%%%%%%%
\begin{figure}
\includegraphics[width=8cm,clip]{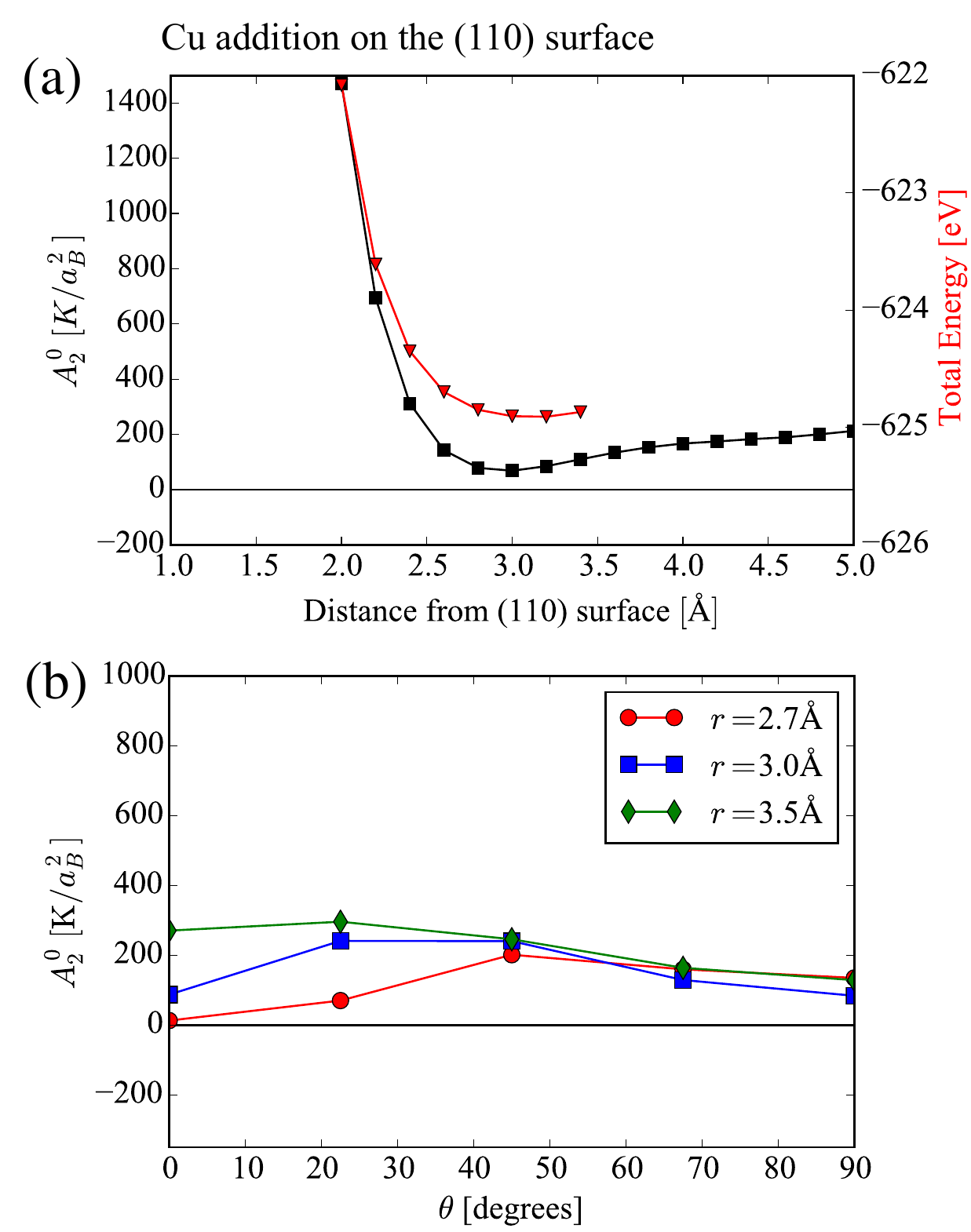}% Here is how to import EPS art
\caption{(a)\,The $r$ dependence of crystal field parameter $A_2^0$ and the total energy for Cu addition on the (110) surface of Nd$_2$Fe$_{14}$B for $\theta=90^\circ$.
The closed squares and triangles indicate $A_2^0$ and total energies, respectively.
(b) The $\theta$ dependence of $A_2^0$ for $r=$ 2.7\,{\AA}, 3.0\,{\AA}, 3.5\,{\AA}.}
\label{f7}
\end{figure}
%%%%%%%%%%%%%%%%%%%%%%%%%%%%%%%%%%%%%%%%%%%%%%%
%
%
Figure \ref{f7}(a) shows the $r$-dependence of the values of $A_2^0$ and electronic total energy for the case of Cu addition to the (110) surface with the angle $\theta=90^\circ$.
Contrary to the case of O addition in Fig.\,\ref{f5}(a), $A_2^0$ maintains positive values for all $r$.
This indicates weak interactions between Nd and Cu atoms.
The steep increase of $A_2^0$ with decreasing $r$ for $r<2.6$\,{\AA} reflects the significance of the nucleus effect of Cu on the 4f electron cloud.
The weak hybridization between Nd and Cu atoms for $r>2.6$\,{\AA} can be seen in the $\theta$-dependence of $A_2^0$, as shown in Fig.\,\ref{f7}(b).

\section{SUMMARY}
Motivated by a recent theoretical work\cite{r10} demonstrating that a surface atomic layer with negative magnetic anisotropy constant $K_u$ can drastically decrease the coercivity $H_c$, we evaluated the influence of trace elements O or Cu on the crystal field parameter $A_2^0$ of the Nd ion, at both the (001) and (110) surface of the Nd$_2$Fe$_{14}$B grain.

In both cases of O and Cu additions to the (001) surface, decreasing the distance $r$ to the surface with constant $\theta=0^\circ$ changes $A_2^0$ from negative to positive values.
At the equilibrium position, the value of $A_2^0$ is found to remain positive.
With further decrease of $r$, $A_2^0$ abruptly becomes negative again with an energy cost less than 1\,eV.
This is due to the positive charge of the nucleus in O and Cu, which gains influence with decreased distance from Nd.
Therefore, the value of $A_2^0$ may exhibit a negative value due to stresses, defects, or deformations around grain boundaries adjacent to the GB phases in an actual system.

The $\theta$-dependence of $A_2^0$ can be roughly expressed as $3\cos^2\theta-1$ for both O and Cu additions.
This suggests that these elements redistribute the valence electrons within the Nd atomic sphere such that the negative point charge model is applicable, as if these species had a negative charge.
%{\color{red}
We observed that the strength of these addition effects is larger with O than with Cu.
This different behavior of O and Cu atoms is clearly seen in the case of (110) surface.
For O-addition, the $r$-dependence of $A_2^0$ is opposite to that in the (001) surface case, as is expected from geometrical effects.
Actually, the surface $K_u$ potentially decreases to negative values at the equilibrium position of O.
However, For Cu-addition in the case of (110) surface, the variation is small compared to O addition, and $A_2^0$ remains positive for all $r$.
%}

Therefore, O is expected to produce negative interfacial $K_u$ more frequently than Cu when it approaches the Nd ion at the grain surface.
The analysis of the total energy showed that local stable positions of the trace element exist for the special configurations considered here.
However, due to the complex interatomic interactions and local stresses in real multi-grain structures of Nd-Fe-B magnets, many possible configurations exist in the local crystalline structure near the interface between GB and Nd$_2$Fe$_{14}$B phases.
In this sense, the ($r$, $\theta$) dependences of the local $K_u$ (i.e., $A_2^0$) shown in this study may apply when we consider the effect of individual atoms adjacent to Nd ions at the interfaces of GBs.

\section*{ACKNOWLEDGMENTS}
This work was supported by CREST-JST.

\begin{thebibliography}{99}%
%
\bibitem{r1}
H.~Kronm\"{u}ller, K.-D.~Durst, and G.~Martinek, J. Magn. Magn. Mater. {\bf 69}, 149 (1987).
%
\bibitem{r2}	
J.~F.~Herbst, Rev. Mod. Phys. {\bf 63}, 819 (1991).
%
\bibitem{r3}
A.~Sakuma, S.~Tanigawa, and M.~Tokunaga, J. Magn. Magn. Mater. {\bf 84}, 52 (1990).
%
\bibitem{r4}
K.~Hono and H.~Sepehri-Amin. Scripta Mater.~{\bf 67}, 530 (2012).
%
\bibitem{r5}
T.~G.~Woodcock, Y.~Zhang, G.~HrKac, G.~Ciuta, N.~M.~Dempsey, T.~Schrefl, O.~Gutfleisch, and D.~Givord,
Scripta Mater. {\bf 67}, 536 (2012).
%
\bibitem{r6}
M.~Sagawa, S.~Hirosawa, H.~Yamamoto, S.~Fujimura, and Y.~Matsuura, Jpn. J. Appl. Phys. {\bf 26}, 785 (1987).
%
\bibitem{r7}
J.~Fidler and K.~G.~Knoch, J. Magn. Magn. Mater. {\bf 80}, 48 (1989).
%
\bibitem{r8}
H.~Sepehri-Amin, T.~Ohkubo, T.~Shima, K.~Hono, Acta Mater. {\bf 60}, 819 (2012).
%
\bibitem{yamada}
M.~Yamada, H.~Kato, H.~Yamamoto and Y.~Nakagawa, Phys. Rev. B {\bf 38}, 620 (1988).
%
\bibitem{a20_1}
M.~Richter, P.~M.~Oppeneer, H.~Eschrig, and B.~Johansson, Phys. Rev. B {\bf 46}, 13919 (1992).
%
\bibitem{a20_2}
M.~F\"{a}hnle, K.~Hummler, M.~Liebs, T.~Beuerle, Appl. Phys. A {\bf 57} 67 (1993).
%
\bibitem{r9}
H.~Moriya, H.~Tsuchiura, and A.~Sakuma, J. Appl. Phys. {\bf 105}, 07A740 (2009).
%
\bibitem{r10}
C.~Mitsumata, H.~Tsuchiura, and A.~Sakuma, Appl. Phys. Express {\bf 4}, 113002 (2011).

\bibitem{paw}
P.~E.~Bl\"{o}chl., Phys. Rev. B, {\bf 50}, 17953 (1994),
 
\bibitem{GGA}
J.~P.~Perdew, J.~A.~Chevary, S.~H.~Vosko, K.~A.~Jackson, M.~R.~Pederson, D.~J.~Singh, and C.~Fiolhais, Phys. Rev. B {\bf 46}, 6671 (1992),
%
\bibitem{vesta}
K.~Momma and F.~Izumi, J. Appl. Crystallogr. {\bf 44}, 1272 (2011).



%
%\bibitem{r11}
%J.~P.~Perdew, J.~A.~Chevary, S.~H.~Vosko, K.~A.~Jackson, M.~R.~Pederson, D.~J.~Singh, and C.~Fiolhais,
%Phys. Rev. B {\bf 46}, 6671 (1992).
%
\bibitem{r12}
T.~Suzuki, Y.~Toga, and A.~Sakuma, J. Appl. Phys. {\bf 115}, 17A703 (2014).
%
\bibitem{r13}
K.~Hummler and M.~F\"{a}hnle, Phys. Rev. B {\bf 53}, 3290 (1996).
%
\bibitem{r13_2}
R.~Skomski, J.~M.~D.~Coey, J. Magn. Magn. Mater. {\bf 140}, 965 (1995).
%
\bibitem{r14}
M.~Yamaguchi and S.~Asano, J. Phys. Soc. Jpn. {\bf 63}, 1071 (1994).
%			
\bibitem{r15}
A.~Sakuma, J. Phys. Soc. Jpn. {\bf 61}, 4119 (1992).
%
\bibitem{r16}
T.~Miyake, K.~Terakura, Y.~Harashima, H.~Kino, and S.~Ishibashi, J. Phys. Soc. Jpn. {\bf 83}, 043702 (2014).
%
%
\end{thebibliography}
\end{document}